%% file: main.tex
\RequirePackage{lineno}

\documentclass[preprint,12pt]{revtex4}

\usepackage{epsfig}
\usepackage{amsmath,amssymb,bm}
\usepackage{hyperref}
\usepackage{placeins}
\usepackage{pstricks}

\usepackage{blindtext}

\begin{document}

\title{Defiducialization: Providing Experimental Measurements for Accurate Fixed-Order Predictions}

\author{A. Glazov}
\email{alexander.glazov@desy.de}
\address{Deutsches Elektronen-Synchrotron DESY, Hamburg, Germany}
\preprint{DESY 20-001}

\begin{abstract}
  An experimental procedure is proposed to perform measurements of differential cross sections for vector boson production
  which can be compared to
  fixed-order QCD predictions with improved accuracy.
  The procedure relies on applying theoretical acceptance corrections  computed as a function of the
  transverse momentum  of the $W/Z$ boson, $p_T$, to the experimental measurement, rather than
  comparing data directly against fiducial fixed-order predictions.
  It is demonstrated that, contrary to standard fiducial computations,
  these acceptance factors vary little at low $p_T$, so they can be reliably computed using fixed-order perturbation theory. 
  An example analysis is performed using the ATLAS measurement of the $Z$-boson
  production cross section at center-of-mass energy of $8$ TeV. The resulting full phase space measurement of the cross section differential in
  the boson rapidity is compared to theoretical predictions computed with next-to-next-to leading-order accuracy in QCD.
  Further extensions of the approach which include different types of measurements and improved theoretical predictions are discussed. 
\end{abstract}


\maketitle

\input{intro}

\input{results}

\input{summary}

\section*{References}
\bibliographystyle{atlasnote}
\bibliography{main}

\end{document}

%% file: intro.tex
\section{Introduction}
\label{sec:intro}

Accurate knowledge of parton distribution functions of the proton (PDFs) is essential for the physics program at the LHC.
PDF uncertainties are the leading source of systematics for precision measurements of the $W$-boson mass and
effective weak mixing angle, $\sin^2\theta_W$~\cite{Aaboud:2017svj,Sirunyan:2018swq}. Reduction of the PDF uncertainties is important in particular for
the interpretation of high statistics measurements for the future LHC data.

PDFs can be constrained using  $W$- and $Z$-boson production in the charged and neutral current Drell-Yan processes.
These processes can be measured with sub-percent experimental accuracy and thus provide a valuable input for the PDF determination.
An example of an accurate measurement is the ATLAS  result on  $\gamma^*/Z$- and $W$-boson production 
at the center-of-mass energy of $\sqrt{s}=7$~TeV~\cite{Aaboud:2016btc}. For the $Z$ boson, the measurement is performed in bins of the invariant
mass of the lepton pair, $m_{\ell\ell}$, and of the lepton pair rapidity, $y_{\ell\ell}$. For the $W$ boson, the results are reported as
a function of the lepton pseudorapidity, $\eta_{\ell}$. Excluding the global normalisation uncertainty, the measurement reaches better than $0.5\%$
experimental uncertainty.

The cross sections differential in $y_{\ell\ell}$, $m_{\ell\ell}$ and $\eta_{\ell}$
are known at next-to-next-to leading-order (NNLO) accuracy in perturbative QCD~\cite{Anastasiou:2003yy,Anastasiou:2003ds,Melnikov:2006di,Catani:2009sm,Catani:2010en}.
For these observables, the corresponding computations are inclusive in the boson transverse
momentum $p_T$ and thus insensitive to $\ln p_T/m_{\ell\ell}$ divergences, providing a robust input for determination of collinear PDFs. 
However the ATLAS measurement is performed in a fiducial volume with experimental cuts on the lepton transverse momentum and lepton pseudorapidity which are
required due to  the detector acceptance. These selection criteria introduce dependence on $p_T$ modelling thereby spoiling accuracy of the fixed-order predictions.
The study performed in Ref.~\cite{Aaboud:2016btc} shows that predictions become unstable
with respect to small variations of the cuts. In addition, calculations obtained using various public NNLO codes 
differ as much as $1\%$ which is significant compared to the experimental accuracy.

The  impact of the fiducial cuts on fixed-order predictions is under investigation since some time. It is known to be large when transverse momentum
of individual leptons (or jets) approaches half of the invariant mass of the lepton pair (or jet pair). It has been proposed in Ref.~\cite{Frixione:1997ks} to introduce
an asymmetry for the cut values for the leading and sub-leading objects.
This procedure seems to stabilize the computation of the fixed-order predictions with respect to small variation of experimental cuts,
however a closer analysis shows that it does not eliminate the uncertainty arising from the logarithmic corrections
which remains similar to the symmetric cuts cases~\cite{Salam?}.
Computation of higher order corrections and/or  inclusion of the $\ln p_T/m_{\ell\ell}$ resummation  should reduce the uncertainty arising from the fiducial cuts.
However the full next-to-next-to-next-to  leading-order  (N$^3$LO)  corrections for Drell-Yan processes are not available yet, while resummation corrections may bring additional uncertainties from the recoil prescription~\cite{Catani:2015vma}.

Experimentally, it is sometimes possible  to isolate regions in the phase space which are not affected by fiducial cuts. For example, in Ref.~\cite{Aaboud:2017ffb} the fiducial acceptance for
the triple differential measurement of $\frac{\mathrm{d}\sigma}{\mathrm{d}y_{\ell\ell} \mathrm{d}\cos \theta \mathrm{d} m_{\ell\ell}}$,
where $\cos\theta$  is the lepton polar angular variable,  is above $99\%$ for  $|\cos\theta|<0.4$, $y<1$ and $m_{\ell\ell}>66$~GeV~\cite{Walker:2019bjd}. 
This region can be safely compared to fixed-order predictions.
Other experimental methods include explicit correction to the full phase space during the data analysis, as in e.g. Ref.~\cite{Aaltonen:2011nr}.

In this paper, another approach is proposed.
It is based on an observation that an acceptance correction from the fiducial to full phase space
has moderate dependence on $p_T$, including low $p_T$ region, as discussed in section~\ref{sec:form}, so that it 
can be determined using fixed-order calculations.
This is demonstrated in section~\ref{sec:valid} using the ATLAS measurement of  $Z$-boson production differential in $y_{\ell\ell}$ and $p_T$ performed using data collected at 
$\sqrt{s}=8$~TeV \cite{Aad:2015auj}. The acceptance correction is computed at NLO for the $Z$ plus jet process ($O(\alpha_S^2)$,
where $\alpha_S$ is the strong coupling constant), using the MCFM v6.8 program~\cite{Campbell:1999ah}, interfaced to APPLGRID~\cite{Carli:2010rw}.
In section~\ref{sec:res}, the correction is used to determine the full phase space measurement differential
in $y_{\ell\ell}$ and $p_T$.
The result is integrated in $p_T$ and the inclusive $\mathrm{d}\sigma/\mathrm{d}y_{\ell\ell}$ differential
cross-section measurement is compared to the  NNLO ($O(\alpha_S^2)$)
computation for inclusive $Z$-boson production obtained using the MCFM v9 program~\cite{Boughezal:2016wmq,Campbell:2019dru}.
To illustrate the importance of the acceptance effects, another comparison is performed in which the fiducial data are integrated in $p_T$ directly
and compared to the fiducial predictions computed to the same order.  While these two comparisons are performed at the same accuracy of the
fixed order predictions in $O(\alpha_S^2)$, a visible difference is observed with the approach advocated in this paper giving a better overall
description of the data.
The paper concludes with a discussion of the results, possible applications of the method
to other measurements, and possible further studies of theoretical predictions.

%% file: results.tex
\section{Formalism}
\label{sec:form}

Recent experimental measurements are usually reported in a fiducial phase space, as determined by the acceptance
of the experimental apparatus. This phase space is defined by selection criteria applied to the final
state objects, such as lepton transverse momenta and pseudorapidities. The presence of experimental cuts must be addressed by the theory predictions.
In the currently adopted procedure, the theoretical predictions are extended to be fully differential, to match the data, so that
they can be also computed in the fiducial phase space.
This procedure is also adapted  for predictions computed using fixed-order calculations. Schematically, the theory to data comparison for $p_T$ integrated observables 
can be written in this case as 
\begin{equation}
  \sigma_{\mathrm{fidu, theory}} = \int \frac{\mathrm{d}\sigma_{\mathrm{full,theory}}}{\mathrm{d}p_T} A(p_T) \mathrm{d}p_T~\mbox{~~~vs~~~} \sigma_{\mathrm{fidu, data}} = \int
  \frac{\mathrm{d}\sigma_{\mathrm{fidu,data}}}{\mathrm{d}p_T} \mathrm{d}p_T  \,. \label{eq:fidu}
\end{equation}
Here a  vector boson transverse meomentum  dependent fiducial acceptance is introduced as $A(p_T) = \sigma_{\mathrm{fidu,theory}}(p_T)/\sigma_{\mathrm{full,theory}}(p_T)$.

Fixed-order predictions for the cross sections  become unreliable at low $p_T$, varying rapidly as $p_T\to 0$. On the other hand, $A(p_T)$ is expected to have only a moderate
dependence on $p_T$, driven by the changes of the boost and polarisation. In general, $A(p_T)$ is not expected to be constant and deviations in  $p_T$ distribution at low $p_T$
may lead to a bias in $\sigma_{\mathrm{fidu, theory}}$.

For experimental measurements which are performed differentially in $p_T$, the comparison between data and theory can be re-arranged such that the data are corrected to the full phase
space first, which is the central proposal of this paper:
\begin{equation}
  \sigma_{\mathrm{full, theory}} = \int \frac{\mathrm{d}\sigma_{\mathrm{full,theory}}}{\mathrm{d}p_T} \mathrm{d}p_T~\mbox{~~~vs~~~} \sigma_{\mathrm{full, data}} = \int
  \frac{\mathrm{d}\sigma_{\mathrm{fidu,data}}}{\mathrm{d}p_T} \frac{1}{A(p_T)} \mathrm{d}p_T \,. \label{eq:full}
\end{equation}
The main advantage of this approach is that the fully integrated in $p_T$
$\sigma_{\mathrm{full, theory}}$ can be computed safely using fixed-order predictions. As it will be illustrated in the following, for  $Z$-boson
production, $A(p_T)$ can be also estimated using fixed-order $Z$ plus jet calculations, including the low $p_T$ region.
The success of this approach relies on smallness of $A(p_T)$ variation at low $p_T$,
which is caused mainly by kinematic effects, while polarisation effects are suppressed at low $p_T$ as $p_T^2/m_Z^2$, where $m_Z$ is the $Z$-boson mass.

To motivate why this would be the
case, it is instructive to decompose the differential cross section in terms
of harmonic polynomials, as discussed, for example, in Ref.~\cite{Gauld:2017tww}.
The fully differential $Z$-boson production cross section can be expressed as
\begin{equation} \label{eq:master} 
  \frac{\mathrm{d^5}\sigma}{\mathrm{d}p_T\mathrm{d}y_{\ell\ell}\mathrm{d}m_{\ell\ell}
    \mathrm{d}\cos\theta \mathrm{d}\phi} = \frac{3}{16\pi} \frac{\mathrm{d^3}\sigma^{U+L}}
       {\mathrm{d}p_T\mathrm{d}y_{\ell\ell}\mathrm{d}m_{\ell\ell} } \sum_{i=0}^8 P_i(\cos\theta, \phi) \,.
\end{equation}
Here $\phi$ is the lepton azimuthal angular variable,
$P_i(\cos\theta, \phi)$ are the nine harmonic polynomials, and $\sigma^{U+L}$ is the unpolarised cross section.
The harmonic polynomials depend on eight angular coefficients $A_i(p_T,y_{\ell\ell},m_{\ell\ell})$ which
define fractions of helicity cross sections with respect to the unpolarised case. Equation~\ref{eq:master} relies
on factorization of the $Z$ boson production and decay processes. In essence, it represents production
of a spin one particle and its following decay.
It is violated by electroweak corrections which
contain interaction of the initial quarks and final state leptons. It is also modified for the $\gamma\gamma \to \ell^+\ell^-$ scattering
process. These corrections are small at the $Z$-pole region and neglected in the following, however it is possible to extend the proposed
procedure and include them.

For the measurements insensitive to the azimuthal angle $\phi$ and forward-backward asymmetry in $\cos\theta$, the most important
coefficient is $A_0$. This is the case for all fiducial measurements of  $Z$-boson production at the LHC, which use charge-symmetric cuts on lepton rapidities.
The coefficient $A_0$ vanishes for $p_T \to 0$ and saturates at $A_0 = 1$ for $p_T \gg 100$~GeV.

To understand better the $p_T$ dependence of the angular coefficients $A_i$, and of the most important coefficient $A_0$ in particular, it is useful
to follow the geometrical approach of Refs.~\cite{Peng:2015spa,Chang:2017kuv}. For the coordinate system in which the $z$-axis is aligned with the momentum direction
of the $q\bar{q}$ pair producing the $Z$ boson, $A_0=0$, as follows from helicity conservation arguments. The direction of the $z$-axis in this coordinate system
depends on radiation and in general does not agree with the $z$-axis of the fixed physical rest frame,
such as given by the direction of the incoming hadrons (laboratory frame).
Defining $\theta_1$ as the angle between the two axes, it can be shown based on simple geometrical arguments that
$A_0$ can be calculated as $A_0=\langle \sin^2 \theta_1 \rangle$
where the average is taken over all event configurations.  Given that $\theta_1 \sim p_T / m_Z$ for small $p_T$, the coefficient $A_0$ is expected to vanish as
$p^2_T/m^2_Z$ for $p_T\to 0$. 

The coefficients $A_i(p_T,y_{\ell\ell},m_{\ell\ell})$ can be calculated using fixed-order
predictions. It is possible that the $A_2$ coefficient is sensitive to non-perturbative effects at low $p_T$~\cite{Chiappetta:1986yg}, however this should have a small impact
on the acceptance. It was demonstrated in Ref.~\cite{Gauld:2017tww} which compared predictions for the $A_i$ coefficients to
the data obtained by the ATLAS and CMS collaborations~\cite{Khachatryan:2015paa,Aad:2016izn}. The calculations were performed
at LO ($O(\alpha_S)$),  NLO ($O(\alpha^2_S)$) and NNLO ($O(\alpha^3_S)$) order for the $Z$-boson plus jet process, as provided by the NNLOJET
program~\cite{Ridder:2015dxa}. For the coefficient $A_0$, good perturbative convergence is observed with already LO predictions being in excellent agreement with the data as shown in Ref.~\cite{Aad:2016izn} for $p_T$ down to
$2.5$~GeV~\footnote{Note that~\cite{Aad:2016izn} counts order of the predictions
  based on inclusive $Z$ production, with NLO corresponding to $O(\alpha_S)$ correction}.

Given the $p_T,y_{\ell\ell},m_{\ell\ell}$ values and the $A_i$ coefficients, the kinematics of the final state leptons and thereby the fraction
of events passing fiducial cuts is fully determined. Any fixed-order prediction for the $Z$ boson plus jet process must obey decomposition of equation~\ref{eq:master}.
Therefore, the fiducial acceptance for each  $p_T,y_{\ell\ell},m_{\ell\ell}$ bin can be determined at fixed order, provided it is narrow enough
to neglect the $p_T$ dependence inside the bin. The residual theoretical uncertainties arising in this procedure can be estimated the usual way, by PDF and
scale variations.

An  estimate of the uncertainty based on the scale variation may not be perfect, especially for the quantities computed as ratios of cross sections. An attempt to improve it
by a partial decorrelation is used in the current analysis, however it may be still incomplete.
It is also possible that $A_i$ coefficients may be affected significantly by the resummation effects, but study of these is beyond the scope of the current paper.
Further studies may bring additional insights on this issue. It should be noted, however, that fixed order analyses usually ignore uncertainties from fiducial acceptance
effects altogether.


\section{Validation}
\label{sec:valid}
\begin{figure}[t]
  \centerline{
    \includegraphics[width=0.8\linewidth]{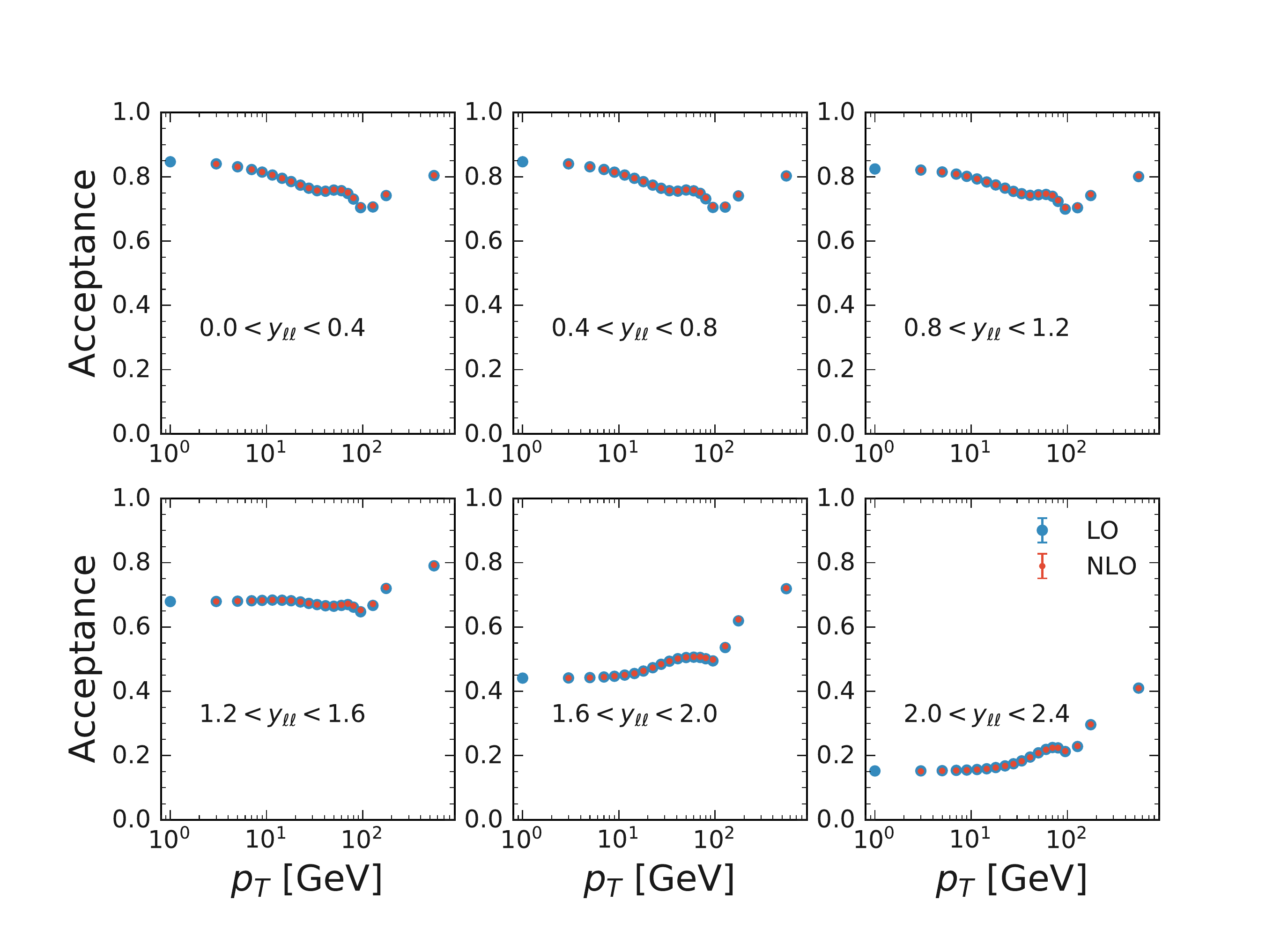}
   }
  \caption{\label{fig:acc}Fiducial acceptance  calculated at LO ($O(\alpha_S)$) and NLO  ($O(\alpha^2_S)$)  using $Z$-boson plus jet process. The error bars
  show total uncertainties of the calculation.}
\end{figure}
The correction procedure is tested using the ATLAS $\gamma^*/Z$ data at $\sqrt{s}=8$~TeV~\cite{Aad:2015auj}. The cross sections
in this analysis are measured differentially in $p_T$ and $y_{\ell\ell}$ for the $66<m_{\ell\ell}<116$~GeV invariant mass range.
There are $20$ variable-size bins in $p_T$, starting with narrow $2$ GeV bins for low $p_T$ and ending with a wide $700$ GeV bin from $200$ to $900$~GeV. 
The $6$ bins in $y_{\ell\ell}$ are equidistant  spanning from $0$ to $2.4$ with a step of $\Delta y_{\ell\ell}=0.4$.
The measurement is performed in the fiducial space defined by the lepton transverse momentum $p_T^{\ell}>20$~GeV and $|\eta_{\ell}|<2.4$ requirements.
All  data tables are taken from the HEPDATA
record of the ATLAS publication. Following Refs.~\cite{Aaboud:2016hhf,Aaboud:2016zpd},
the measurement is re-scaled by the original and updated luminosity ratio: $20.3/20.2$. The total normalization uncertainty of the measurement is $1.9\%$.

\subsection{Estimation of $A(p_T)$ at LO and NLO}

The fiducial acceptance is estimated using the MCFM v6.8 program  interfaced to \mbox{APPLGRID}, for fast evaluation of theoretical uncertainties.
The $Z$ plus jet process is computed at LO 
and NLO yielding acceptance estimates $A_{\mathrm{LO}}$ and
$A_{\mathrm{NLO}}$, respectively. It is also possible to use NNLO calculations for the $Z$ plus jet process which became available recently~\cite{Ridder:2016rzm}, however they are not used in the
present analysis. Note that in order of $\alpha_S$, the NLO calculations for the $Z$ plus jet process match NNLO  calculations for inclusive $Z$-boson production.

The renormalization and factorization scales are set to $m_{\ell\ell}$ and the CT14NNLO PDF set~\cite{Dulat:2015mca} is used for the calculations.
The jet transverse momentum is required to be above $1$~GeV. The APPLGRID is binned in the $Z$-boson transverse momentum and rapidity, following the binning used in the ATLAS measurement.
Note that while at NLO jet and $Z$-boson transverse momenta do not agree, the $1$~GeV jet transvserse momentum cut
has no impact for the  bins  with the $Z$-boson $p_T \ge 2$~GeV.

The resulting fiducial acceptance is shown in Figure~\ref{fig:acc}.
The statistical uncertainties for the LO calculations are negligible and for NLO they are below $0.1\%$ for the majority of the bins. To achieve this accuracy,
computational resources of about $20000$ CPU hours are required.

The fiducial  acceptance is very similar for the three central rapidity bins with $y<1.2$ and it is always above 70\%. For the forward rapidity, the acceptance starts to decrease and drops
below $20\%$ for the $2.0<y_{\ell\ell}<2.4$ bin at low $p_T$. The $p_T$ dependence of the acceptance is also different for forward compared to central rapidities.
The first bin in $p_T$ spans between $0$ and $2$~GeV and NLO calculations of the acceptance become unstable, thus
only the LO result is presented. For all other bins, LO and NLO calculations agree very well.

\begin{figure}
  \centerline{
    \includegraphics[width=0.8\linewidth]{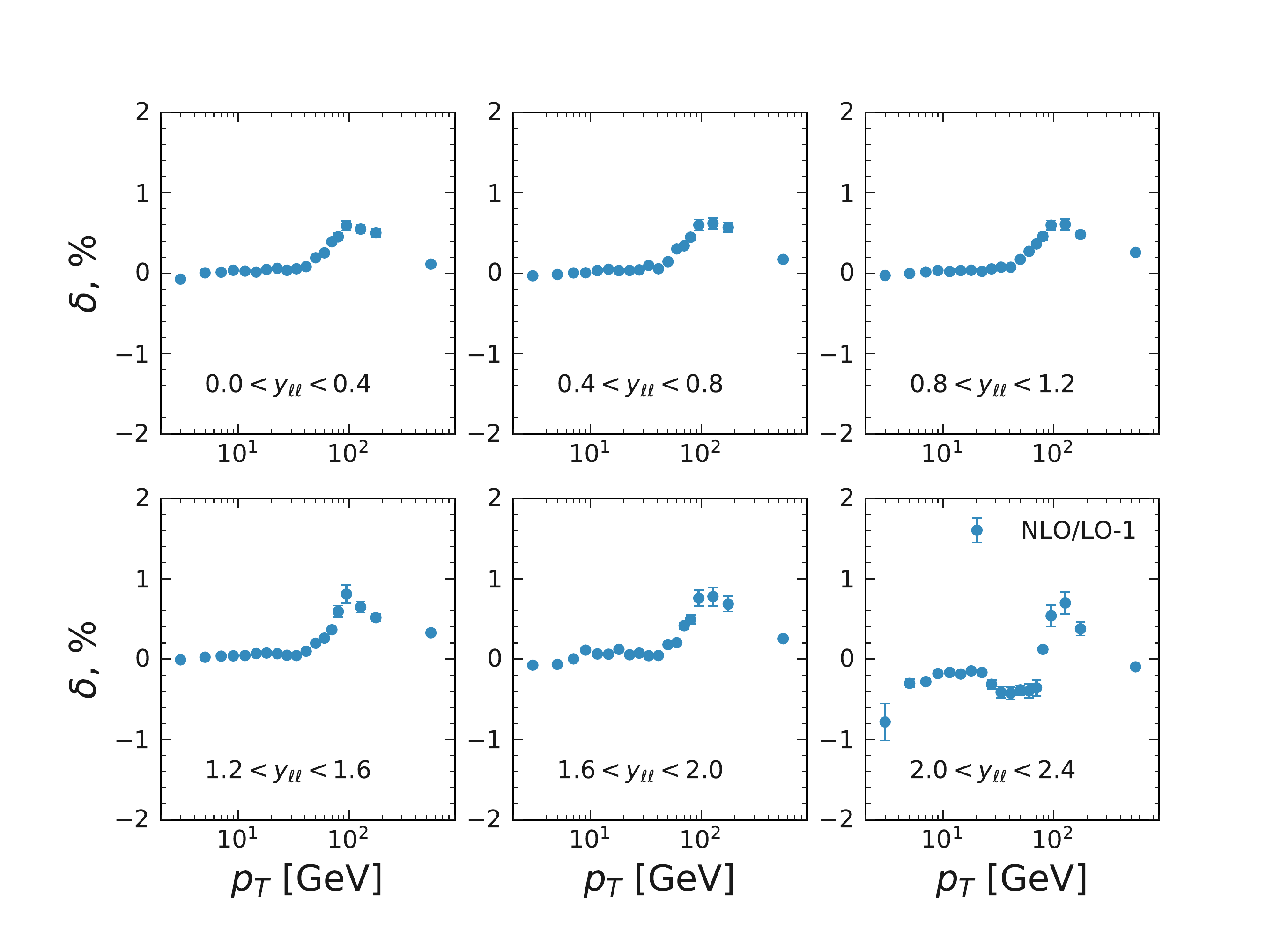}
   }
  \caption{\label{fig:accDelta}Percentage difference between acceptance corrections calculated at NLO and LO. The error bars indicate total uncertainties of the NLO calculations.}
\end{figure}
Figure~\ref{fig:accDelta} presents quantitative comparison of the acceptances computed at different orders in terms of $\delta = A_{\mathrm{NLO}}/A_{\mathrm{LO}}-1$.
The NLO correction is below $0.1\%$ for $p_T<m_Z$ for all rapidity bins with $y_{\ell\ell}<2.0$. For all bins the correction does not exceed $1\%$,
indicating good convergence of the perturbative series. Based on this good agreement and moderate dependence of the correction on $p_T$, the correction for the bin
$0<p_T<2$~GeV is estimated using LO calculation, re-scaled by the ratio of NLO to LO for the bin $2<p_T<4$~GeV.

Moderate variation of the estimated fiducial acceptance as a function of $p_T$ and very good agreement of predictions computed at
LO and NLO provide good confidence that fixed-order predictions can be safely used. Additional support comes from the 
studies of the angular coefficient $A_0$ at NNLO~\cite{Ridder:2015dxa}, mentioned above,
which indicate good perturbative convergence. Note, however, that good perturbative convergence does not necessarily lead to small theoretical uncertainties and furhter
in-depth theoretical studies are required to fully validate this proposal.
With these observations, the analysis proceeds to an estimate of the uncertainties for the NLO-based predictions.
\subsection{Estimation of uncertainties}
\begin{figure}
  \centerline{
    \includegraphics[width=0.8\linewidth]{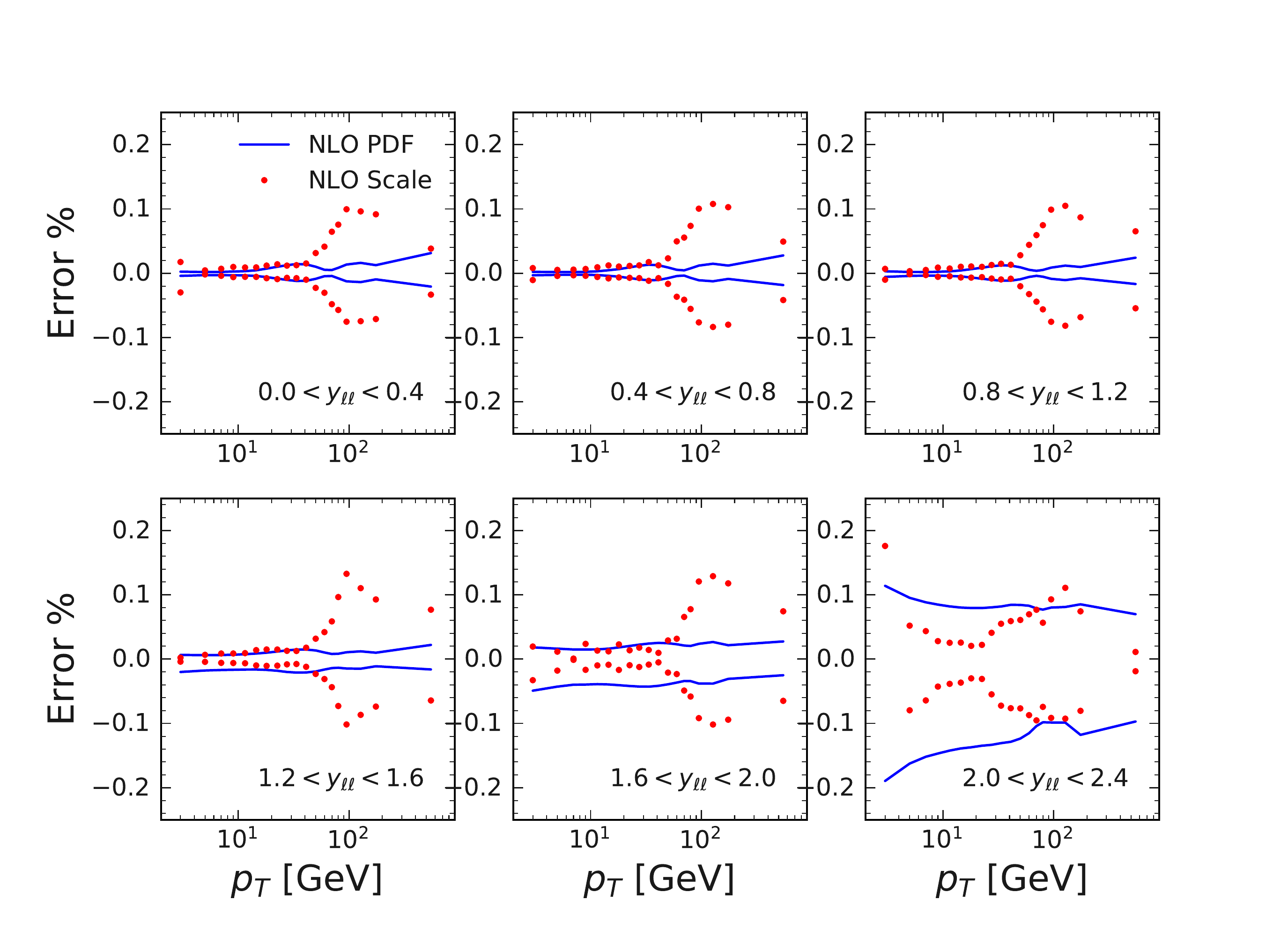}
   }
  \caption{\label{fig:accErr}PDF and scale variation uncertainties of the acceptance correction calculated at NLO.}
\end{figure}
PDF and scale uncertainties of the acceptance computed at NLO are shown in Figure~\ref{fig:accErr}. The PDF uncertainties are evaluated using the CT18ANNLO set~\cite{Hou:2019qau,Hou:2019efy} scaled to $68\%$ c.l. 
The scale uncertainties are estimated by varying factorization and renormalization scale by factor of two and taking the envelope of resulting acceptance factors.
Both PDF and scale uncertainties evaluated this way are typically below $0.1\%$. The PDF uncertainties increase to about $0.15\%$ for the highest rapidity bin.
As an additional check of the PDF dependence, the acceptance is calculated using CT14NNLO,  MMHT14NNLO~\cite{Harland-Lang:2014zoa}, NNPDF31~\cite{Ball:2017nwa},
ABMP16~\cite{Alekhin:2017kpj} and ATLASepWZ16~\cite{Aaboud:2016btc} PDF sets. For all sets except ATLASepWZ16,
the calculated acceptance is found to be  within the the PDF error bands
of the CT18ANNLO set. The ATLASepWZ16 based acceptance agrees with the other sets for $y_{\ell\ell}<2$. For the highest rapidity bin, the acceptance based on ATLASepwz16 deviates
by as much as $-0.2\%$  for low $p_T$ and up to $+0.35\%$ for $p_T$ around  $m_Z$. No additional uncertainty is introduced due to this deviation.

\begin{figure}
  \centerline{
    \includegraphics[width=0.7\linewidth]{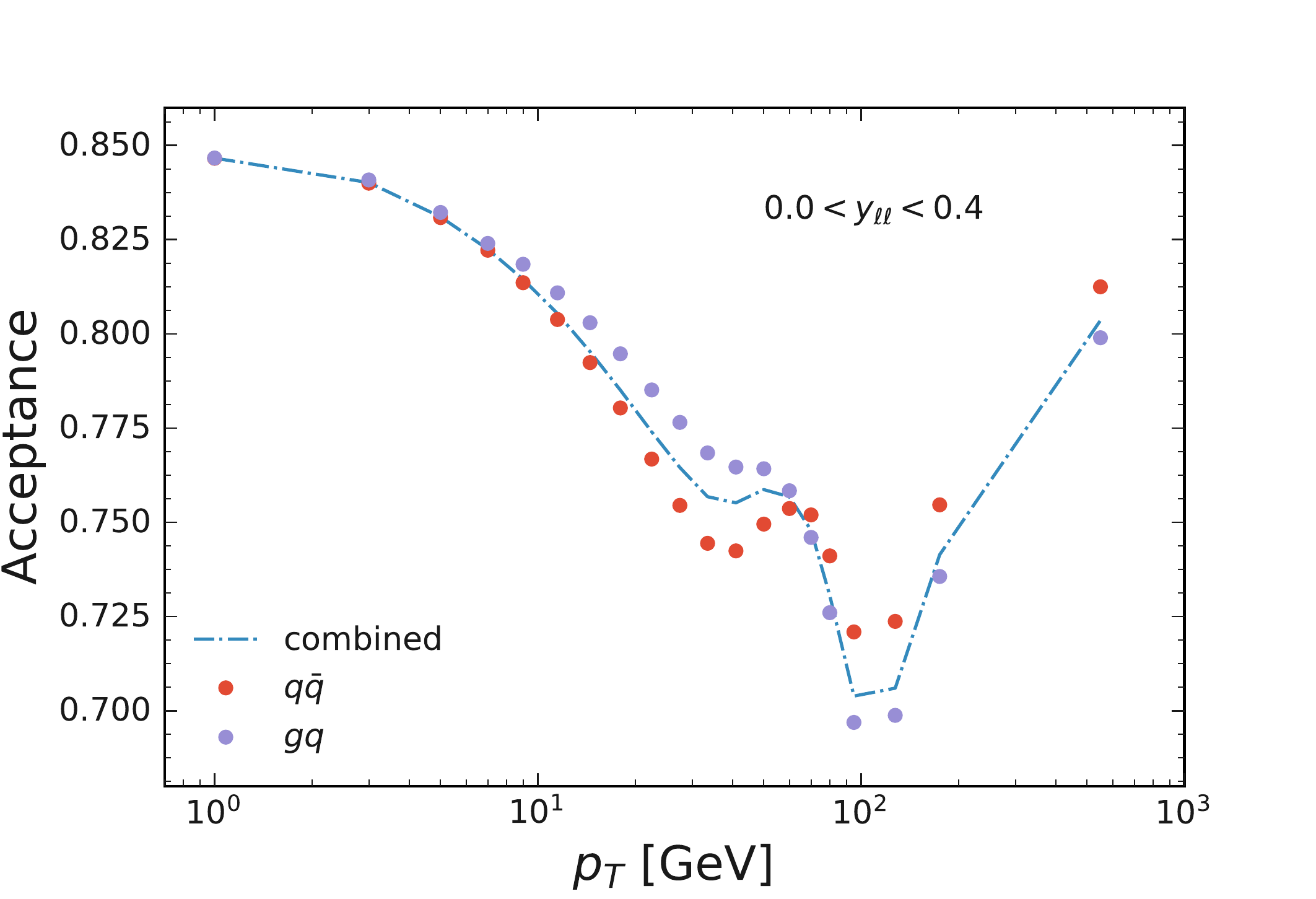}
   }
  \caption{\label{fig:aaqg}Fiducial acceptance calculated for $q\bar{q}$ and $qg$ contributions for $0<y_{\ell\ell}<0.4$.}
\end{figure}
\begin{figure}
  \centerline{
    \includegraphics[width=0.8\linewidth]{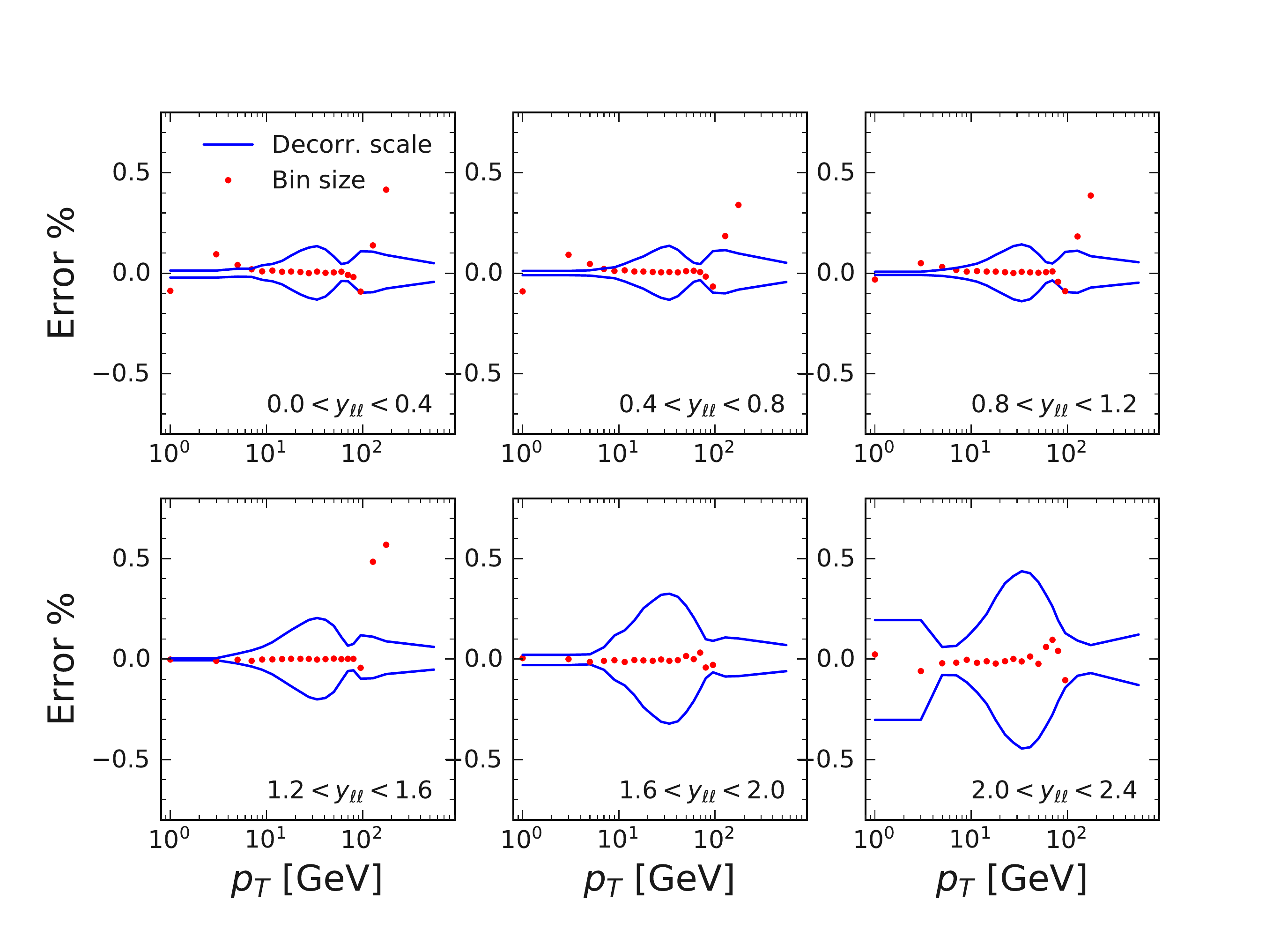}
   }
  \caption{\label{fig:accErr2}Bin size and QCD scale uncertainties calculated assuming independent variations of $q\bar{q}$ and $qg$ contributions.}
\end{figure}

The smallness of the uncertainties is related to the fact that the main effect is purely kinematic variation of the fiducial acceptance as a function of $p_T$ and theoretical
uncertainties arise only from modeling of the polarisation.
To gain some insight on polarisation uncertainties, one can again follow the intuitive geometrical approach of
Refs.~\cite{Peng:2015spa,Chang:2017kuv}.
For the predictions at LO, the contributions from $q\bar{q}$ and $qg$ subprocesses lead to different average mis-alignment angle $\theta_1$ yielding in different
$A_0$. It is shown in Refs.~\cite{Peng:2015spa,Chang:2017kuv} that the polarisation coefficients observed in data do indeed lie between the two predictions. 
APPLGRID based calculations allow to separate contributions of $q\bar{q}$ and $qg$ subprocesses,
fiducial acceptances for which can be compared to the full result. The resulting   comparison 
is shown in Figure~\ref{fig:aaqg} for $|y|<0.4$.
Scale variations yield only to
small changes of the acceptances of the
individual contributions, and a common scale variation for both of them yields to a moderate change  in their relative size.
This may explain the smallness
of the acceptance uncertainty due to the scale variation seen in Figure~\ref{fig:accErr}.
It is possible that higher order corrections
may have different impact on $q\bar{q}$ and $qg$ subprocesses, and to be conservative the scale variations are considered as independent for them
with the total scale uncertainty evaluated as a sum in quadrature.
The resulting uncertainty is shown in Figure~\ref{fig:accErr2} is always larger compared to the standard
approach shown in Figure~\ref{fig:accErr}.

As it is mentioned above, the difference in $p_T$ distribution between the data and predictions should not matter for fiducial acceptance estimation, as long as bin sizes are small enough such that variation of the acceptance within the bin can be neglected. The bins are however rather large in this analysis
compared to variations of the acceptance at low $p_T$, see Figure~\ref{fig:acc}. There is also a large difference between data and fixed order predictions at low $p_T$.
The effect of acceptance variation within the bin is studied at LO. Predictions are generated from $p_T>0.1$~GeV, $p_T>10$~GeV, and $p_T>100$~GeV with $0.1$~GeV, $1$~GeV, and $20$~GeV bin size, respectively. For these bin sizes the variation of the fiducial acceptance can be neglected since it is much smaller compared to the data statistical uncertainty for the corresponding enclosing analysis bins.
The full phase-space $p_T$ distribution is then reweighted to follow one estimated in the data. The latter is approximated by first correcting the measured result to the full phase space
using acceptance as in  Figure~\ref{fig:acc} and then fitting by a phenomenological function $f(p_T) = p_T \sum_{i=1}^{N} a_i \exp(- p_T b_i)$ with
$N=3$ and $a_i, b_i$ treated as free parameters. The fits are performed in each $y$ bin independently.
The function describes the data well for the whole $p_T$ range. Its shape is very different from the fixed order predictions for $p_T=0$~GeV with $f(0\,\mathrm GeV{})=0$.
The reweighted full-phase space distribution is corrected for the fiducial acceptance in $0.1,1,20$~GeV $p_T$ bins. The resulting fiducial and full phase-space distributions are
integrated following the analysis binning and the binned fiducial acceptance is determined from their ratio. For the lowest $p_T<2$~GeV bin, the fixed order distributions
are integrated starting form $1$~GeV, to emulate the analysis procedure, while the reweighted distributions are integrated from $0$~GeV with fiducial acceptance for $p_T<0.1$~GeV
estimated using a linear extrapolation. 

The difference between acceptances determined using reweighted and original LO $p_T$ distribution is taken as systematic uncertainty which  is shown in Figure~\ref{fig:accErr2}.
The uncertainty reaches about $0.1\%$ for the two lowest $p_T$ bins for $y<0.4$, where $p_T$ variation of the acceptance is significant.
The uncertainty for the lowest $p_T$ bin is reduced by using $p_T>1$~GeV for the fixed order predictions.
For $p_T>100$~GeV, the uncertainty starts to increase and reaches few percent level, however it remains smaller than the statistical uncertainty of the data.

To summarise the study of uncertainties, the largest effect arises from the
decorrelated scale variation (up to $0.5\%$ for $p_T \sim m_Z/2$ ) followed by PDFs (up to $0.15\%$ for $y_{\ell\ell}>2$).
The bin-size effects remain below $0.1\%$ for the bulk of the distribution. It should be noted, however, as it was already discussed, that
the current analysis ignores the effect of electroweak corrections and potential resummation effects.
\section{Application and Results}
\label{sec:res}
\begin{figure}
  \centerline{
    \includegraphics[width=0.7\linewidth]{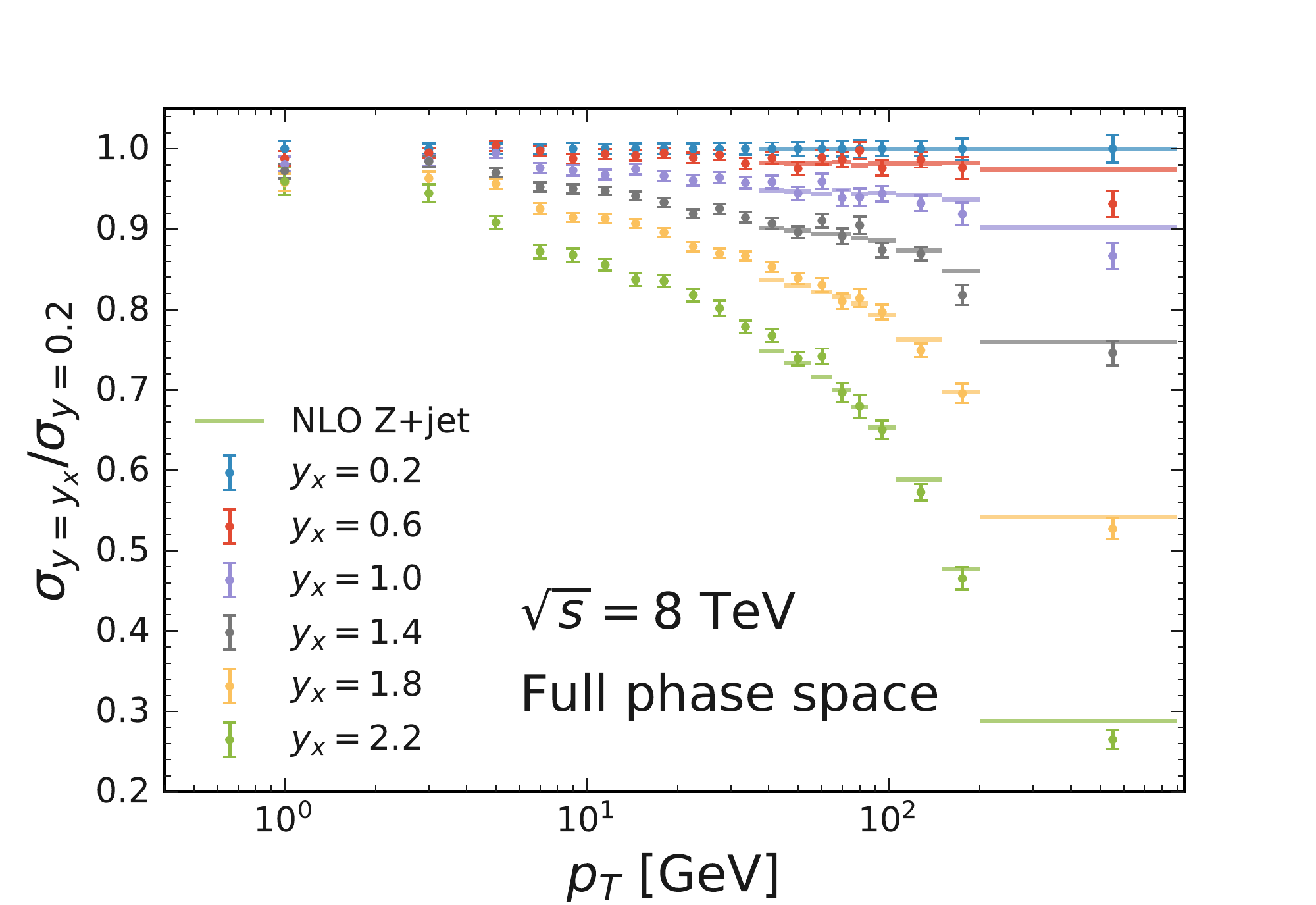}
  }
  \caption{\label{fig:ptFull} Ratios of differential $\mathrm{d}\sigma/\mathrm{d}p^{\ell\ell}_T$ full phase space distributions for different
    $y_{\ell\ell}$ bins to the $0<y_{\ell\ell}<0.4$ bin. The error bars show total uncertainties of individual $\mathrm{d}\sigma/\mathrm{d}p^{\ell\ell}_T$ measurements.
    The horizontal lines shown for $p_T>35$~GeV represent NLO predictions for $Z$ plus jet process using CT14NNLO PDF set.
  }
\end{figure}
With the caveat discussed above that the estimate may be not complete,
the size of the theoretical uncertainties on the acceptance corrections suggests that it is possible to proceed with applying them to the data. Figure~\ref{fig:ptFull} shows
the ratio of the $\mathrm{d}\sigma/\mathrm{d}p_T$ distributions corrected to the full phase space for a given $y_{\ell\ell}$ bin to the most central $0<y_{\ell\ell}<0.4$ bin. 
The ratio is close to one for small $p_T$ decreasing almost linearly for large $p_T$ and large $y_{\ell\ell}$. For $p_T>35$~GeV, the data are compared to fixed-order predictions at NLO 
for the $Z$-boson plus jet process computed using the same APPLGRID as  for the acceptance correction which is convoluted with the CT14NNLO PDF set. The data and predictions are found to be in  reasonable agreement.

\begin{figure}
  \centerline{
    \includegraphics[width=0.7\linewidth]{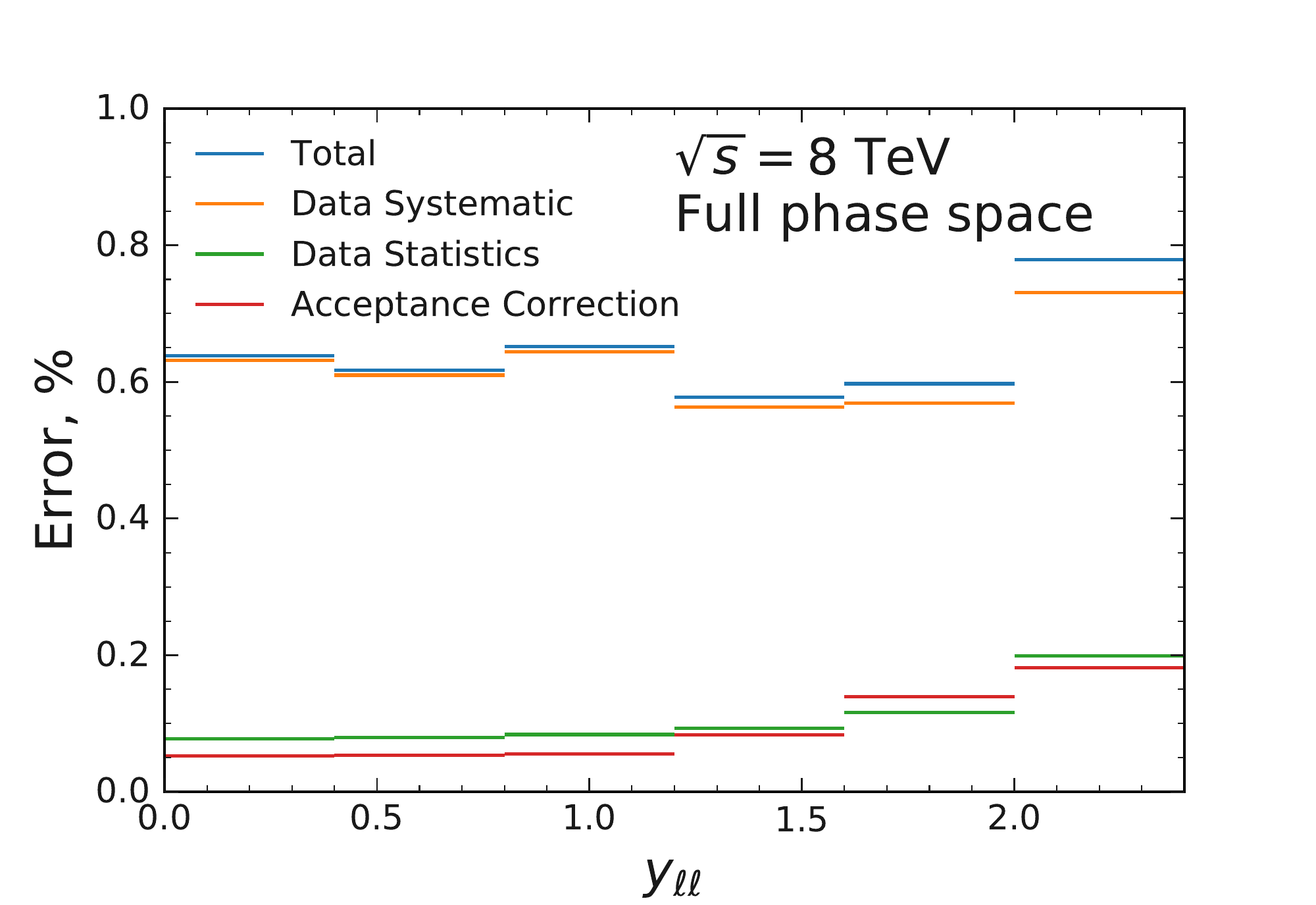}
  }
  \caption{\label{fig:erryz} Decomposition of the uncertainties for the $\mathrm{d}\sigma/\mathrm{d}y_{\ell\ell}$ measurement. The global normalization uncertainty due to the luminosity
  measurement of $1.9\%$ is not shown.}
\end{figure}
As the next step the data are integrated in $p_T$. The correlated uncertainties are propagated linearly while the statistical uncertainties are combined in quadrature.
The uncertainty decomposition is shown in figure~\ref{fig:erryz}. The uncertainties are dominated by the correlated experimental errors. The statistical uncertainties
are at around $0.1\%$ except the highest rapidity bin. The acceptance correction uncertainties are comparable to the statistical uncertainties.

\begin{figure}
  \centerline{
    \includegraphics[width=0.7\linewidth]{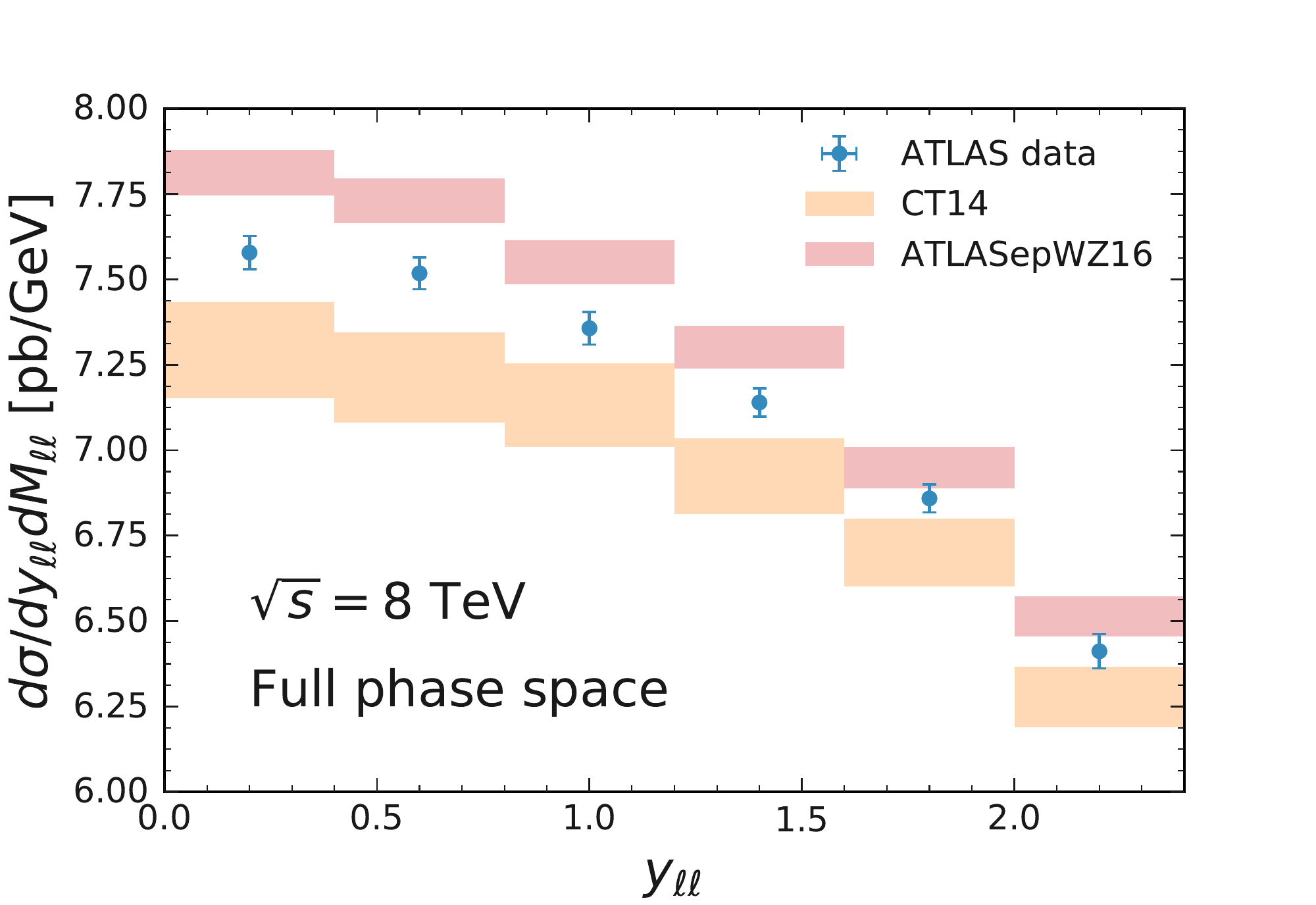}
  }
  \caption{\label{fig:yz} Full phase space measurement of $\mathrm{d}\sigma/\mathrm{d}y_{\ell\ell}$ compared to predictions using various PDFs computed at NNLO using MCFM v9.
    Global normalization uncertainty  of $1.9\%$ is not shown.  Error bars show total data uncertainties.
    Bands indicate PDF uncertainties of the predictions.  For ATLASepWZ16, the uncertainties are computed using the ATLASepWZ16-EIG set only. 
}
\end{figure}
The measurements are compared to NNLO predictions obtained using the MCFM v9 program. Two PDF sets are used for the comparison: CT14NNLO and ATLASepWZ16.
CT14NNLO is representative of other global PDF sets with reduced compared to $\bar{u}$ and $\bar{d}$  strange-quark distribution.
ATLASepWZ16, on the other hand, is obtained by fitting the ATLAS fiducial $W,Z$ data at $\sqrt{s}=7$~TeV from Ref.~\cite{Aaboud:2016btc} and has enhanced strangeness.
The MCFM program is used with nominal settings, electroweak corrections disabled, and scales set to $m_{\ell\ell}$. 
The central value of the total cross section for $\sqrt{s}=8$~TeV and $66<m_{\ell\ell}<116$~GeV  
obtained for the CT14NNLO PDF set is $1114.9(1)$~pb, which is $0.4\%$ larger than the value of $1110(1)$~pb
obtained in Ref.~\cite{Aaboud:2016zpd} using the DYTURBO program~\cite{Camarda:2019zyx}.

The comparison between the full phase space measurement of $\mathrm{d}\sigma/\mathrm{d}y_{\ell\ell}$ at $\sqrt{s}=8$~TeV and predictions is shown in figure~\ref{fig:yz}.
The global normalization uncertainty of $1.9\%$ is omitted from the figure and the predictions are shown with PDF uncertainties only. The uncertainties are larger
for the CT14NNLO  compared to ATLASepWZ16-EIG PDF set due to usage of increased tolerance factors compared to the $\Delta \chi^2=1$ criterion adapted for the  ATLASepWZ16-EIG set.
The predictions based on CT14NNLO (ATLASepWZ16) underestimate (overestimate) the data. Given the significant global normalization uncertainty the difference is however not 
significant.

\begin{figure}
  \centerline{
    \includegraphics[width=0.7\linewidth]{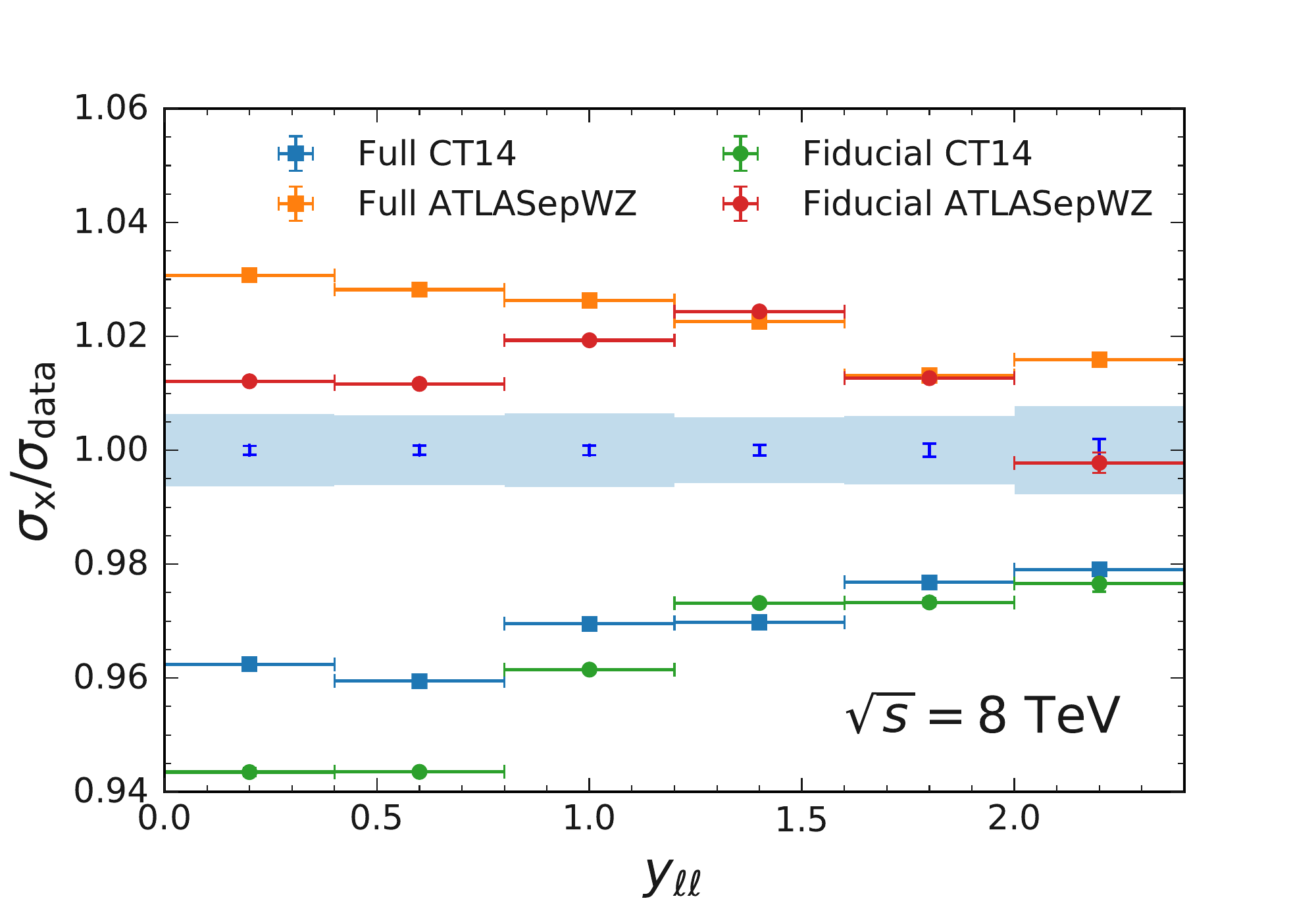}
  }
  \caption{\label{fig:fullfidu}
    Ratio of the predicted to the measured $\mathrm{d}\sigma/\mathrm{d}y_{\ell\ell}$  cross section.
    The ratios are calculated in the full or fiducial phase space using the CT14NNLO and ATLAS-epWZ PDF sets, as indicated in the legend.
    The error band (bars)
    centered at one shows the data total experimental (statistical) uncertainty. The error bars for the ratios of the data to predicted cross sections show estimated statistical uncertainties
    of the predictions.  Global normalization uncertainty  of $1.9\%$ is not shown.
  }
\end{figure}
More quantitatively, the comparison can be performed by taking ratios of the predictions to the data. It is also interesting to compare the ratios of the full cross sections to the ratios
of the fiducial cross sections, to see the impact of the $p_T$-dependent acceptance corrections. The fiducial cross section in data is computed by simple integration over the
differential cross section. It is verified that the total fiducial cross section agrees  with the one reported in the ATLAS publication. The NNLO predictions are obtained using the MCFM v9
program with fiducial cuts enabled.  The resulting full and fiducial cross-section ratios for the two PDF sets are shown in figure~\ref{fig:fullfidu}.
For the CT14NNLO based prediction, the ratio of the full phase space cross sections is closer to the unity compared to the ratio of the fiducial cross sections.
The dependence of the ratio on $y_{\ell\ell}$ is also reduced significantly. For the ATLASepWZ16 based prediction, the fiducial ratio is closer to unity, which is not too surprising since the set
is fitted to the fiducial ATLAS measurement at $\sqrt{s}=7$~TeV.  The difference between full and fiducial ratios is similar for both predictions for the five $y_{\ell\ell}<2.0$ bins.
For the first three bins the fiducial ratio is lower than the full one, by as much as $2\%$, for the two lowest $y$ bins. For bins with $1.2<y_{\ell\ell}<2.0$, they roughly agree. For the highest
rapidity bin, the behavior is different depending on the PDF set. For the CT14NNLO based prediction the two ratios agree while for the ATLASepWZ16 based prediction the fiducial ratio
is below the full one by almost $2\%$.

For the full phase space comparison based on the CT14NNLO and ATLASepWZ16 sets,
the ratios of $\sigma_{\mathrm{theory}}/\sigma_{\mathrm{data}}$ show opposite trend as a function of $y_{\ell\ell}$. Given that the strange-quark 
distribution affects low rapidity region more, based on this observation, it is possible to make a conjecture that the ATLAS data may prefer somewhat larger strangeness content than
in the CT14NNLO set and somewhat smaller than in the ATLASepWZ16 set. The  overshoot of the predictions  based on the ATLASepWZ16 set is likely to be caused by the
bias in the fiducial NNLO prediction which was used in the analysis of the ATLAS data at $\sqrt{s}=7$~TeV.

\begin{figure}[tb]
  \centerline{
    \includegraphics[width=0.7\linewidth]{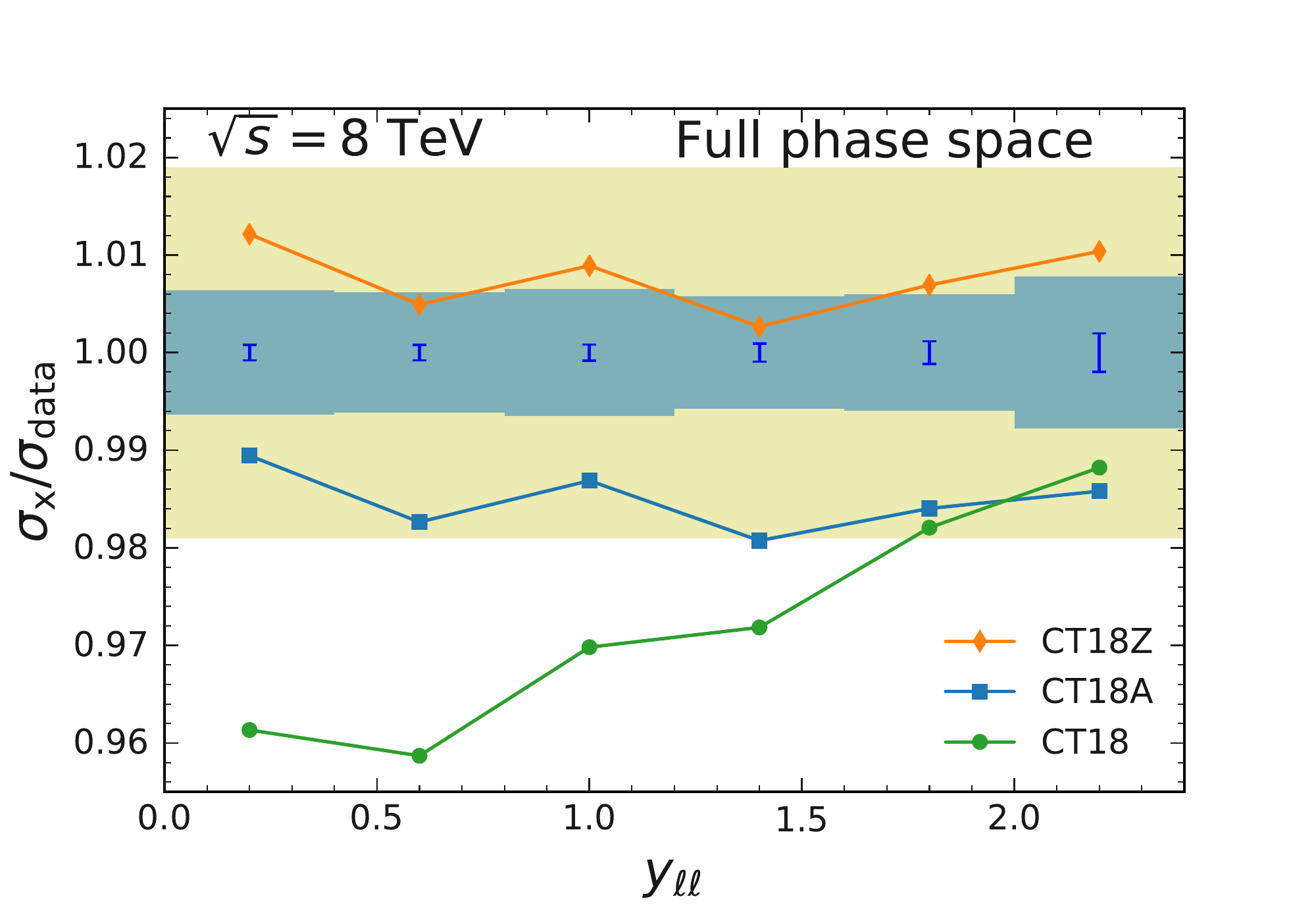}
  }
  \caption{\label{fig:fullct18}
    Ratio of the predicted to the measured $\mathrm{d}\sigma/\mathrm{d}y_{\ell\ell}$  cross section.
    The ratios are calculated in the full phase space using the CT18, CT18A and CT18Z NNLO PDF sets, as indicated in the legend.
    The error bands and error bars    centered at one show the data normalisation, total experimental (excluding normalisation) and statistical uncertainty.  
  }
\end{figure}

To check this, predictions based on CT18NNLO, CT18ANNLO and CT18ZNNLO PDF sets~\cite{Hou:2019qau,Hou:2019efy}  are compared to the data in the full phase space.
The results are shown in figure~\ref{fig:fullct18}.
Compared to the CT18NNLO set, both CT18ANNLO and CT18ZNNLO  sets include the ATLAS $W,Z$ data at $\sqrt{s}=7$~TeV.
As a result, the CT18ANNLO and CT18ZNNLO sets have an enhanced strange-quark distribution, with no suppression compared to other light quarks for Bjorken $x<0.01$.
The enhancement is however smaller compared to the ATLASepWZ16 set since the CT18 sets use other data which favor smaller strangeness. 
The strange-quark distribution in CT18NNLO is similar to CT14NNLO.
The CT18ANNLO set in addition uses a charm-quark mass that is increased compared to other PDF  sets.
This leads
to  suppression of the charm-quark contribution, compensated by an increase of the light-quark sea, yielding in turn an increase of the Drell-Yan
cross sections at the LHC, as it was demonstrated in Ref.~\cite{Abramowicz:1900rp}. The predictions based on the CT18ANNLO and CT18ZNNLO sets show better agreement
with the ATLAS data compared to those based on the  CT18NNLO set:  the ratio shows no dependence on $y_{\ell\ell}$ and the overall normalisation of the ratio is closer to unity. 

\section{Discussion}
In the discussion above, it is argued that for comparison of 
fixed-order predictions with data, it is better 
to use full (equation~\ref{eq:full})
instead  of  fiducial (equation~\ref{eq:fidu}) cross sections.
Both formula require accurate prediction for the acceptance $A(p_T)$, however equation~\ref{eq:full} does not require accurate modeling of the $p_T$ distribution by the prediction.

The accuracy of the fixed order predictions for $A(p_T)$ may be improved further by using existing higher order calculations.
Furthermore, electroweak corrections should be included as well. Comparison of $A(p_T)$ computed using pure fixed order predictions and ones which include
resummation effects can be used to quantify importance of the latter for the data analysis. 
If needed, the PDF uncertainty can be reduced further, by using the same PDF in the calculation of $\sigma_{\mathrm{full,theory}}$ and $A(p_T)$.  
This could be relevant for PDF fits in particular and can be arranged without difficulty since
the PDF-dependent acceptance  $A(p_T)$ can be computed using exact and fast methods such as APPLGRID. 

The method is applicable to the measurements differential in the boson transverse momentum $p_T$. It is natural for $Z$-boson production, but can be extended for the $W$ boson too.
Given that $A(p_T)$ has only mild variation at low $p_T$ (see figure~\ref{fig:acc}), it is probably possible to use coarser $p_T$ binning, which is required
by the experimental resolution. Since the rapidity of the $W$ boson can not be reconstructed, this may lead to increased PDF dependence of the acceptance correction factor, requiring
its update when performing comparisons for predictions based on different PDFs. Studies of the defiducialization method for $W$-boson production are however beyond the scope of the current
analysis and they await for experimental measurements, such as of the lepton pseudorapidity distribution, reported in bins of the $W$-boson transverse momentum.

%% file: summary.tex
\section{Summary}
In summary, a method is proposed to perform comparisons of experimental data on Drell-Yan production, performed in fiducial phase space,
and fixed-order QCD predictions with improved accuracy.
The method requires the data to be measured differentially in the boson transverse momentum
$p_T$ in addition to other variables of interest such as $y_{\ell\ell}$ and $m_{\ell\ell}$.
It relies on the ability to calculate the fiducial acceptance correction
for a given $p_T$ value at fixed order. The method should be extended in future by inclusion of electroweak corrections and estimation of resummation effects.

The method is applied to the ATLAS $Z$-boson production data at a center-of-mass energy of $8$~TeV.
The data are integrated in $p_T$ and compared to the NNLO predictions. To check impact of the new approach, the comparison is also performed using fiducial
cross sections. A significant, compared to experimental uncertainties, improvement in the data description by the predictions based on CT14NNLO set is observed when performing comparison
in the full phase space.

\section{Acknowledgements}
The author would like to thank Simone Amoroso, Ludovica Aperio Bella, Maarten Boonenkamp, 
Stefano Camarda, and Jan Kretzschmar for discussions and comments to the paper draft.